\documentclass[12pt]{article}
\usepackage{epsfig,amssymb,amsmath}

\usepackage{graphicx}

\newcommand{\eba}{\begin{array}}
\newcommand{\eea}{\end{array}}
\newcommand{\ebe}{\begin{eqnarray}}
\newcommand{\eee}{\end{eqnarray}}
\newcommand{\eb}{\begin{equation}}
\newcommand{\ee}{\end{equation}}

\renewcommand\Pi{{{P}}}

\renewcommand\eb{\begin{equation}}
\renewcommand\ee{\end{equation}}

\renewcommand\Pi{{{P}}}

\begin{document}
\noindent {\bf A strategy on prion AGAAAAGA amyloid fibril molecular modeling}\\ [0.4cm] 
Jiapu Zhang\\ [0.2cm]
School of Sciences, Information Technology and Engineering, University of Ballarat,\\ 
Mount Helen, VIC 3350, Australia, j.zhang@ballarat.edu.au, Ph: (61)423487360\\ [0.4cm]
{\bf Abstract:} 
X-ray crystallography and nuclear magnetic resonance (NMR) spectroscopy are two powerful tools to determine the protein 3D structure. However, not all proteins can be successfully crystallized, particularly for membrane proteins. Although NMR spectroscopy is indeed very powerful in determining the 3D structures of membrane proteins, same as X-ray crystallography, it is still very time-consuming and expensive. Under many circumstances, due to the noncrystalline and insoluble nature of some proteins, X-ray and NMR cannot be used at all. Computational approaches, however, allow us to obtain a description of the protein 3D structure at a submicroscopic level.

To the best of the author's knowledge, there is little structural data available to date on the AGAAAAGA palindrome in the hydrophobic region (113--120) of prion proteins, which falls just within the N-terminal unstructured region (1--123) of prion proteins. Many experimental studies have shown that the AGAAAAGA region has amyloid fibril forming properties and plays an important role in prion diseases. However, due to the noncrystalline and insoluble nature of the amyloid fibril, little structural data on the AGAAAAGA is available. This paper introduces a simple molecular modeling strategy to address the 3D atomic-resolution structure of prion AGAAAAGA amyloid fibrils. Atomic-resolution structures of prion AGAAAAGA amyloid fibrils got in this paper are useful for the drive to find treatments for prion diseases in the field of medicinal chemistry.\\

\noindent {\bf Keywords } Prion AGAAAAGA palindrome, amyloid fibril, molecular modeling, prion dieseases.

\section{Introduction}
Prion diseases are invariably fatal and highly infectious neurodegenerative diseases affecting humans and animals. The neurodegenerative diseases such as Creutzfeldt-Jakob disease (CJD), variant Creutzfeldt-Jakob diseases (vCJD), Gerstmann-Straussler-Scheinker syndrome (GSS), Fatal Familial Insomnia (FFI), Kuru in humans, scrapie in sheep, bovine spongiform encephalopathy (BSE or mad-cow disease) and chronic wasting disease (CWD) in cattle belong to prion diseases. By now there have not been some effective therapeutic approaches or medications to treat all these prion diseases.\\

Prion diseases are amyloid fibril diseases. The normal cellular prion protein (PrP$^C$) is rich in $\alpha$-helices but the infectious prions (PrP$^{Sc}$) are rich in $\beta$-sheets amyloid fibrils. The conversion of PrP$^C$ to PrP$^{Sc}$ is believed to involve a conformational change from a predominantly $\alpha$-helical protein (42\% $\alpha$-helix, 3\% $\beta$-sheet) to a protein rich in $\beta$-sheets (30\% $\alpha$-helix, 43\% $\beta$-sheet) \cite{griffith1967}.\\

Many experimental studies such as \cite{brown2000, brown2001, brown1994, holscher1998, jobling2001, jobling1999, kuwata2003, norstrom2005, wegner2002} have shown two points: (1) the hydrophobic region (113-120) AGAAAAGA of prion proteins is critical in the conversion from a soluble PrP$^C$ into an insoluble PrP$^{Sc}$ fibrillar form; and (2) normal AGAAAAGA is an inhibitor of prion diseases. Furthermore, we computationally clarified that prion AGAAAAGA segment indeed has an amyloid fibril forming property \cite{zhang2009,zhang2011,zhangsun2011}. However, laboratory experiences have shown that using traditional experimental methods is very difficult to obtain atomic-resolution structures of AGAAAAGA due to the noncrystalline and insoluble nature of the amyloid fibril \cite{tsai2005, zheng2006}. By introducing novel mathematical canonical dual formulations and computational approaches, in this paper we may construct atomic-resolution molecular structures for prion (113–-120) AGAAAAGA amyloid fibrils.\\

Many studies have indicated that computational approaches or introducing novel mathematical formulations and physical concepts into molecular biology can significantly stimulate the development of biological and medical science. Various computer computational approaches were used to address the problems related to ``amyloid fibril" \cite{carter1998,chou2004b, chou2004c,chou2002,wang2008,wei2005}. Here, we would like to use the simulated annealing evolutionary computations to build the optimal atomic-resolution amyloid fibril models in hopes to be used for controlling prion diseases.\\

The atomic structures of all amyloid fibrils revealed steric zippers, with strong van der Waals (vdw) interactions between $\beta$-sheets and hydrogen bonds (HBs) to maintain the $\beta$-strands \cite{sawaya2007}. The vdw contacts of atoms are described by the Lennard-Jones (LJ) potential energy:  
\begin{equation} \label{LJ_r_form}
V_{LJ}(r)=4\varepsilon \left[ (\frac{\sigma}{r})^{12} - (\frac{\sigma}{r})^6 \right] ,
\end{equation}
where $\varepsilon$ is the depth of the potential well and $\sigma$ is the atom diameter; these parameters can be fitted to reproduce experimental data or deduced from results of accurate quantum chemistry calculations. The $(\frac{\sigma}{r})^{12}$ term describes repulsion and the $(\frac{\sigma}{r})^6$ term describes attraction. If we introduce the coordinates of the atoms whose number is denoted by $N$ and let $\varepsilon = \sigma =1$ be the reduced units, the form (\ref{LJ_r_form}) becomes
\begin{equation}\label{LJ_x_form}
f(x)=4\sum_{i=1}^N \sum_{j=1,j<i}^N \left( \frac{1}{\tau_{ij}^6}
-\frac{1}{\tau_{ij}^3} \right),
\end{equation}
\noindent where $\tau_{ij}=(x_{3i-2}-x_{3j-2})^2
                          +(x_{3i-1}-x_{3j-1})^2
                          +(x_{3i}  -x_{3j}  )^2$,
$(x_{3i-2},x_{3i-1},x_{3i})$ is the coordinates of atom $i$, $N \geq 2$. The minimization of LJ potential $f(x)$ on $\mathbb{R}^n$ (where $n=3N$) is an optimization problem: 
\begin{equation} \label{min_LJ_x_form}
\min f(x) \quad subject \quad to \quad x\in \mathbb{R}^{3N}.
\end{equation}
Similarly as (\ref{LJ_r_form}), i.e. the potential energy for the vdw interactions between $\beta$-sheets:
\begin{equation} \label{LJ_AB_form}
V_{LJ}(r)=\frac{A}{r^{12}} -\frac{B}{r^6},
\end{equation}
the potential energy for the HBs between the $\beta$-strands has the formula 
\begin{equation} \label{HB_r_form}
V_{HB}(r)= \frac{C}{r^{12}} -\frac{D}{r^{10}} ,
\end{equation}
where $A,B,C,D$ are given constants. Thus, the amyloid fibril molecular modeling problem is deduced into well solve the optimization problem (\ref{min_LJ_x_form}).\\

This paper is organized as follows. In Section 2, we first describe how to build the prion AGAAAAGA amyloid fibril molecular models, and then explain how the models with 6 variables only are built and can be solved by any optimization algorithm. At the end of Section 2 the models are done a little refinement by Amber 11 \cite{case_etal}. At last, we conclude that when using the time-consuming and costly X-ray crystallography or NMR spectroscopy we still cannot determine the protein 3D structure, we may introduce computational approaches or novel mathematical formulations and physical concepts into molecular biology to study molecular structures. This concluding remark will be made in the last section.  

\section{Prion AGAAAAGA amyloid fibril models' Molecular Modeling and Optimizing}
Constructions of the AGAAAAGA amyloid fibril molecular structures of prion 113--120 region are based on the most recently released experimental molecular structures of human M129 prion peptide 127--132 (PDB entry 3NHC released into Protein Data Bank (http://www.rcsb.org) on 04-AUG-2010). The atomic-resolution structure of this peptide is a steric zipper, with strong vdw interactions between $\beta$-sheets and HBs to maintain the $\beta$-strands (Figure 1).\\ 

In Figure 1 we see that G (H) chains (i.e. $\beta$-sheet 2) of 3NHC.pdb can be obtained from A (B) chains (i.e. $\beta$-sheet 1) by
\begin{equation}
G (H) = \left( \begin{array}{ccc}
 1  &0  &0  \\
 0  &-1 &0  \\
 0  &0  &-1 \end{array} \right) A (B) + 
\left( \begin{array}{c}
 9.07500\\
 4.77650\\
 0.00000 
\end{array} \right),
\end{equation}
and other chains can be got by
\begin{equation}
I (J) = I_3 G (H) +\left( \begin{array}{c} 
0\\
9.5530\\ 
0\end{array} \right),
K (L) = I_3 G (H) + \left( \begin{array}{c} 
0\\
-9.5530\\ 
0\end{array} \right),
\end{equation}
\begin{equation}
C (D)=I_3 A (B)+ \left( \begin{array}{c}
0\\
9.5530\\ 
0\end{array} \right),
E (F) =I_3 A (B) +\left( \begin{array}{c}
0\\
-9.5530\\ 
0\end{array} \right),
\end{equation}
where $I_3$ is the 3-by-3 identity matrix. Basing on the template 3NHC.pdb from the Protein Data Bank, three prion AGAAAAGA palindrome amyloid fibril models –- an AAAAGA model (Model 1), a GAAAAG model (Model 2), and an AAAAGA model (Model 3) –- will be successfully constructed in this paper. AB chains of Models 1-3 were respectively got from AB chains of 3NHC.pdb using the mutate module of the free package Swiss-PdbViewer (SPDBV Version 4.01) ({\small http://spdbv.vital-it.ch}). It is pleasant to see that almost all the hydrogen bonds are still kept after the mutations; thus we just need to consider the vdw contacts only. Making mutations for GH chains of 3NHC.pdb, we can get the GH chains of Models 1-3. However, the vdw contacts between A chain and G chain, between B chain and H chain are too far at this moment (Figure 2).\\

Seeing Figure 2, we may know that for Models 1-3 at least two vdw interactions between A.ALA3.CB-G.ALA4.CB, B.ALA4.CB-H.ALA3.CB should be maintained. Fixing the coordinates of A.ALA3.CB and B.ALA4.CB, letting the coordinates of G.ALA4.CB and H.ALA3.CB be variables, we may get a simple LJ potential energy minimization problem (\ref{min_LJ_x_form}) just with six variables. For solving this six variable optimization problem, {\it any} optimization computational algorithm can be used to solve this low-dimensional problem; for example, in this paper we may use the hybrid discrete gradient simulated annealing method \cite{zhangsun2011}. Setting the coordinates of G.ALA4.CB and H.ALA3.CB as initial solutions, running the hybrid discrete gradient simulated annealing optimization algorithm, for Models 1-3 we get
\begin{equation}
G (H) = \left( \begin{array}{ccc} \label{close_vdw}
 1  &0  &0  \\
 0  &-1 &0  \\
 0  &0  &-1 \end{array} \right) A (B) + 
\left( \begin{array}{c}
 -0.703968\\
 7.43502\\
-0.33248 
\end{array} \right).
\end{equation}
By (\ref{close_vdw}) we can get close vdw contacts between A chain and G chain, between B chain and H chain (Figure 3).\\

Furthermore, we may employ the Amber 11 package \cite{case_etal} to slightly optimize Models 1-3 and at last get Models 1-3 with stable total potential energies (Figure 4). The other CDIJ and EFKL chains can be got by parallelizing ABGH chains in the use of mathematical formulas (7)-(8).

\section{Conclusion}
X-ray crystallography is a powerful tool to determine the protein 3D structure. However, it is time-consuming and expensive, and not all proteins can be successfully crystallized, particularly for membrane proteins. Although NMR spectroscopy is indeed a very powerful tool in determining the 3D structures of membrane proteins, it is also time-consuming and costly. Due to the noncrystalline and insoluble nature of the amyloid fibril, little structural data on the prion AGAAAAGA segment is available. Under these circumstances, the novel simple strategy introduced in this paper can well do the molecular modeling of prion AGAAAAGA amyloid fibrils. This indicated that computational approaches or introducing novel mathematical formulations and physical concepts into molecular biology can significantly stimulate the development of biological and medical science. The optimal atomic-resolution structures of prion AGAAAAGA amyloid fibils presented in this paper are useful for the drive to find treatments for prion diseases in the field of medicinal chemistry.\\

{\small
\noindent {\bf Acknowledgments:} This research was supported by a Victorian Life Sciences Computation Initiative (VLSCI) grant number VR0063 on its Peak Computing Facility at the University of Melbourne, an initiative of the Victorian Government.
}

\begin{figure}[h!]
\centerline{
\includegraphics[scale=1.3]{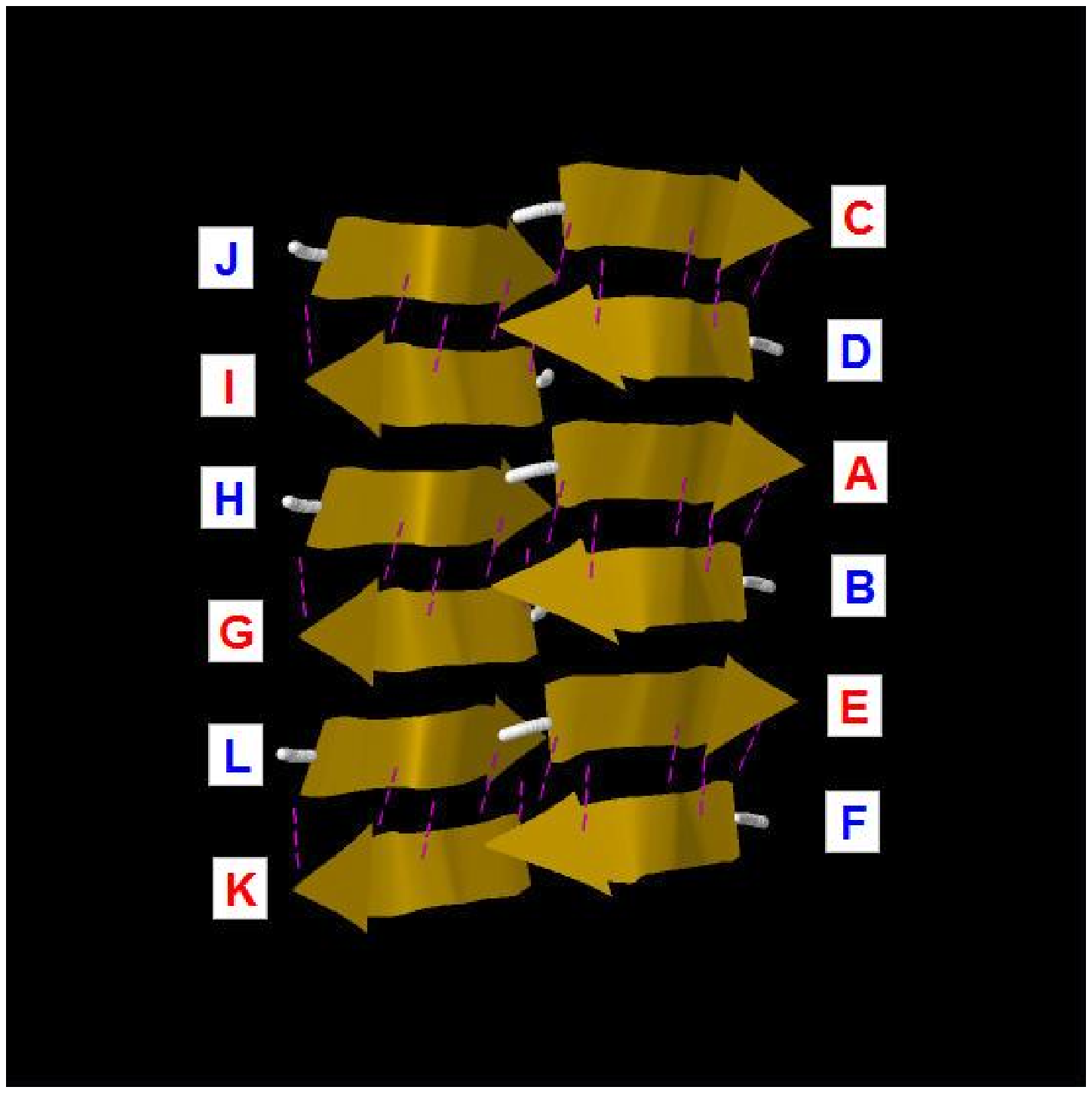}
}
\caption{Protein fibril structure of human M129 prion GYMLGS (127--132). The purple dashed lines denote the hydrogen bonds. A, B, ..., K, L denote the 12 chains of the fibril.}
\label{fig1}
\end{figure}

\begin{figure*}[h!]
\centerline{
\includegraphics[scale=0.6]{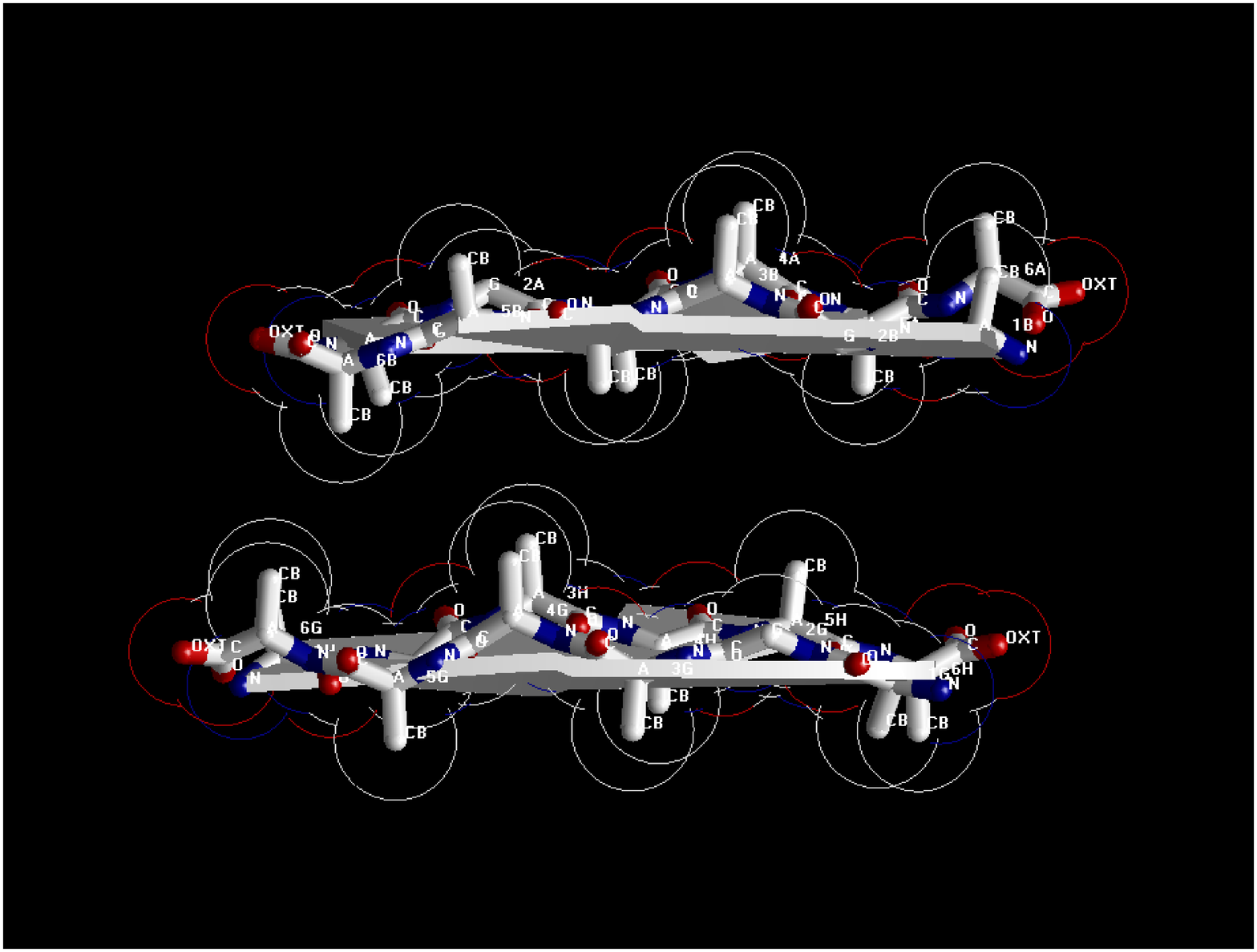}
}
\end{figure*}
\begin{figure*}[h!]
\centerline{
\includegraphics[scale=0.6]{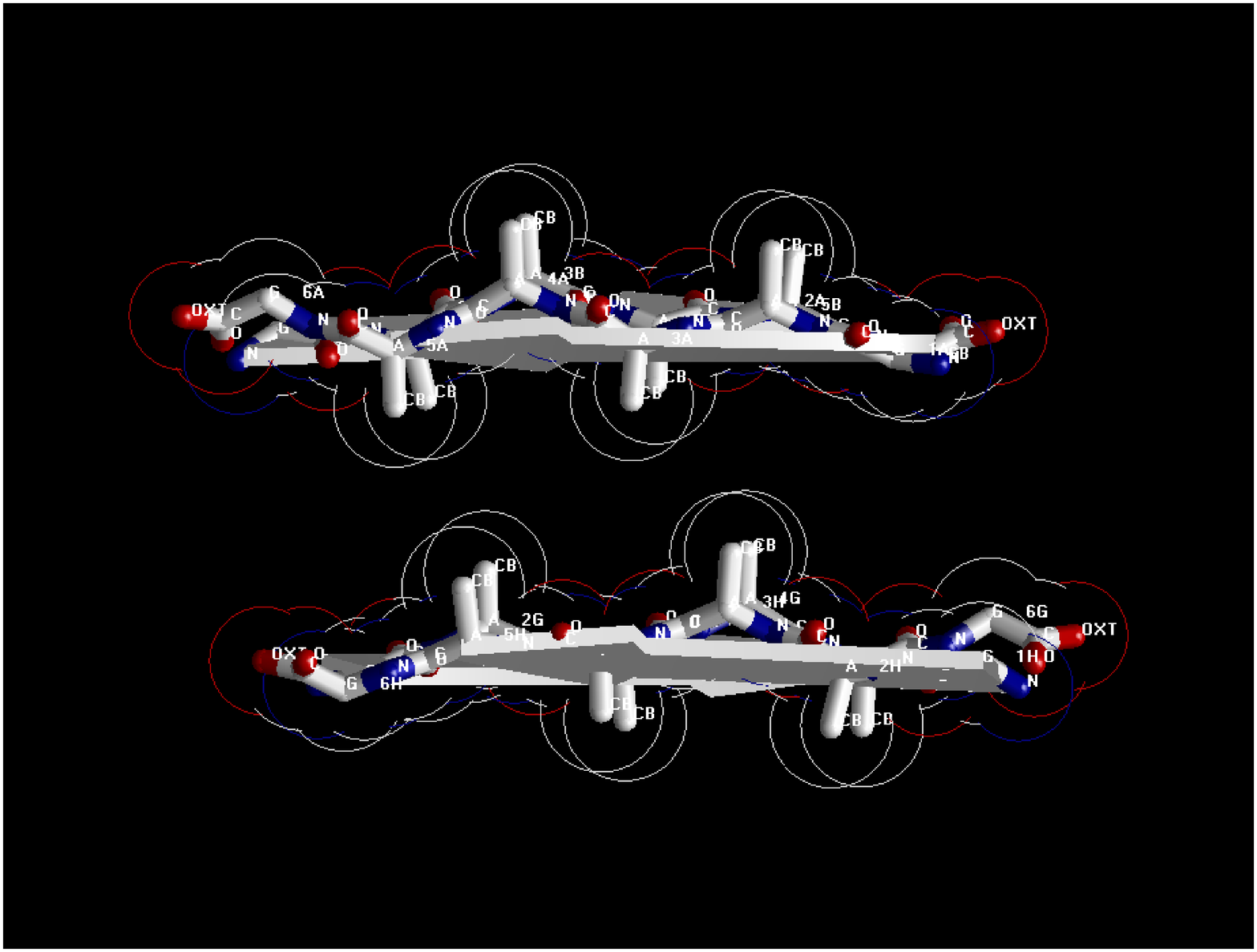}
}
\end{figure*}
\begin{figure}[h!]
\centerline{
\includegraphics[scale=0.6]{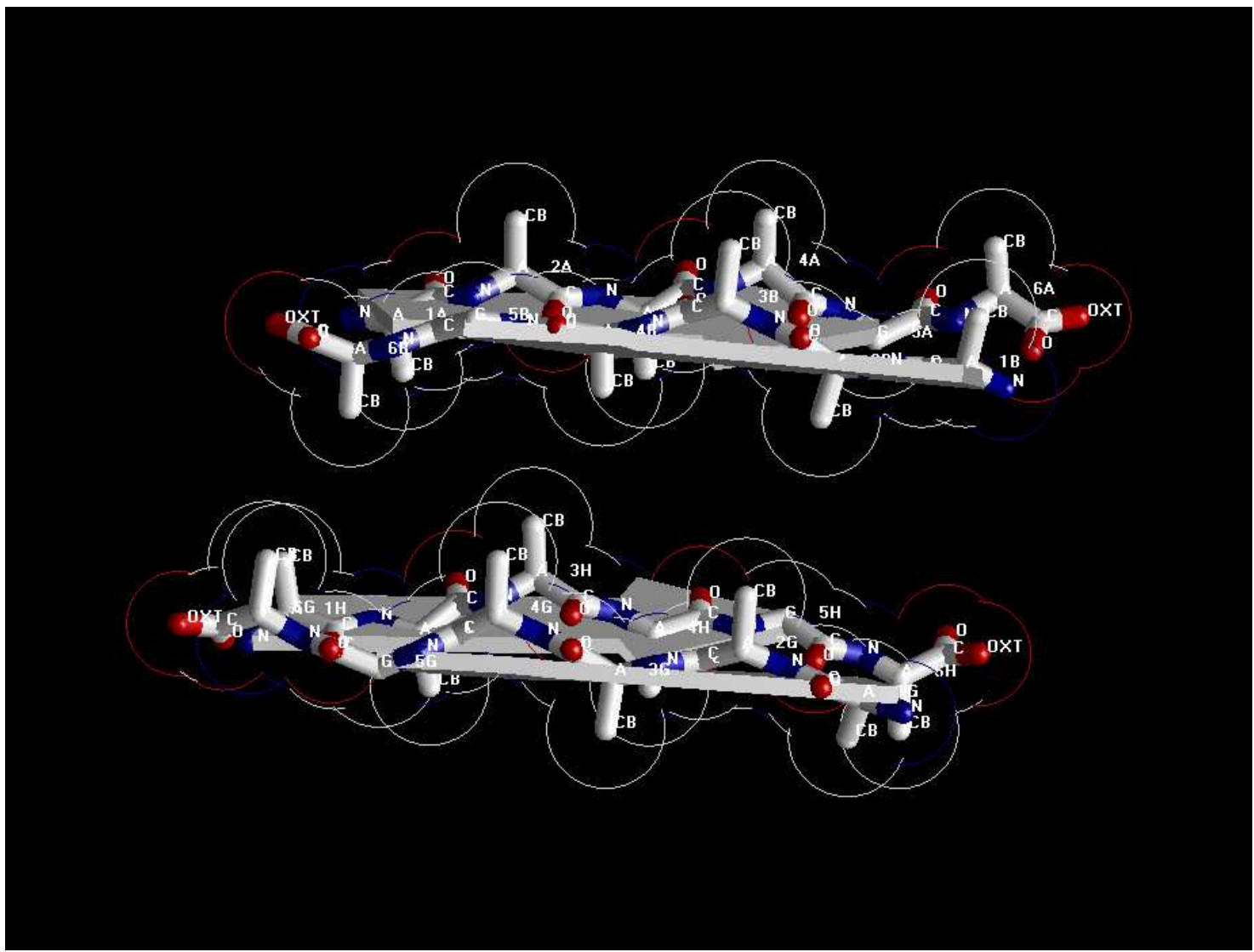}
}
\caption{Far vdw contacts of AG chains and BH chains of Models 1-3.}
\label{fig2_bad_vdw}
\end{figure}

\begin{figure*}[h!]
\centerline{
\includegraphics[scale=0.6]{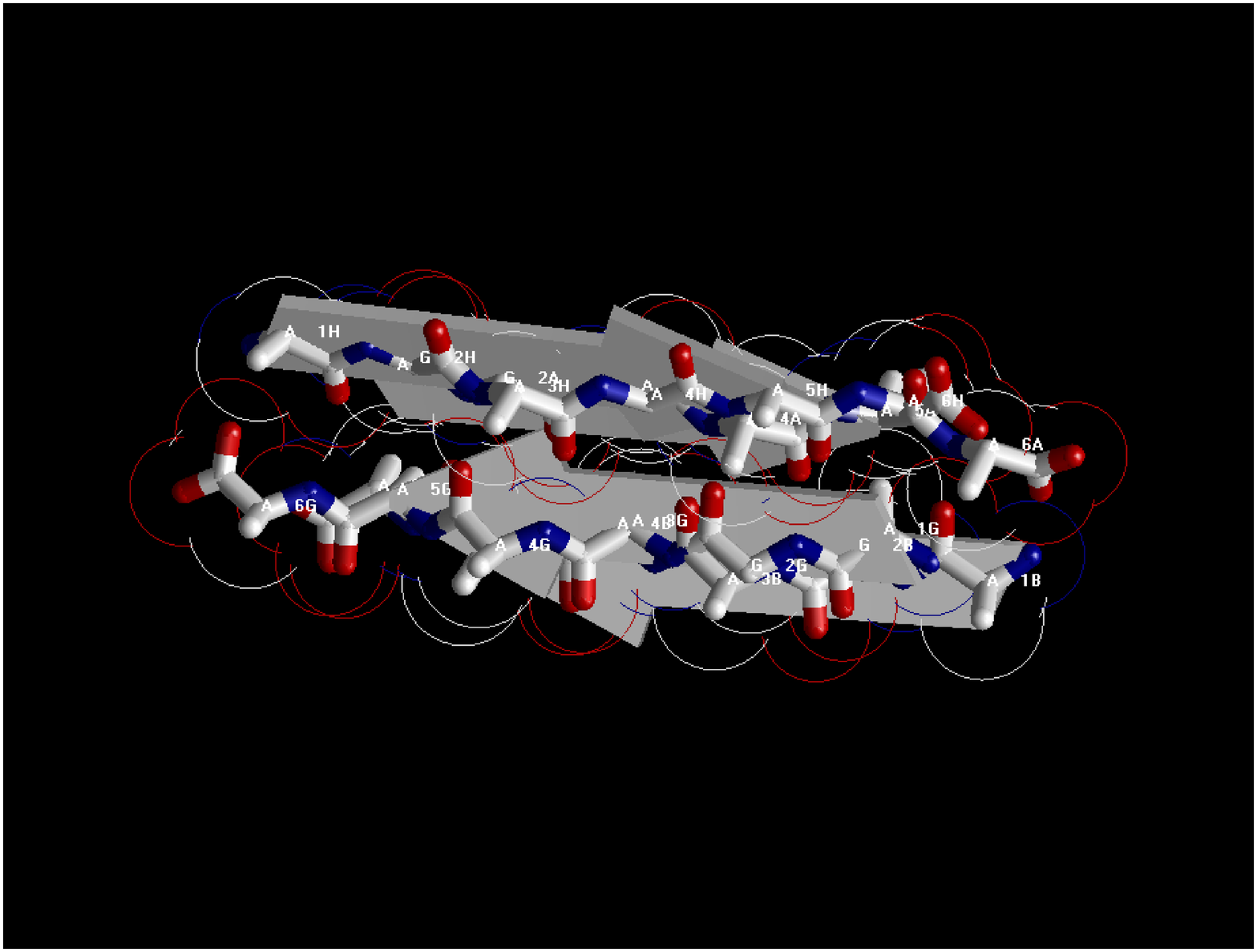}
}
\end{figure*}
\begin{figure*}[h!]
\centerline{
\includegraphics[scale=0.6]{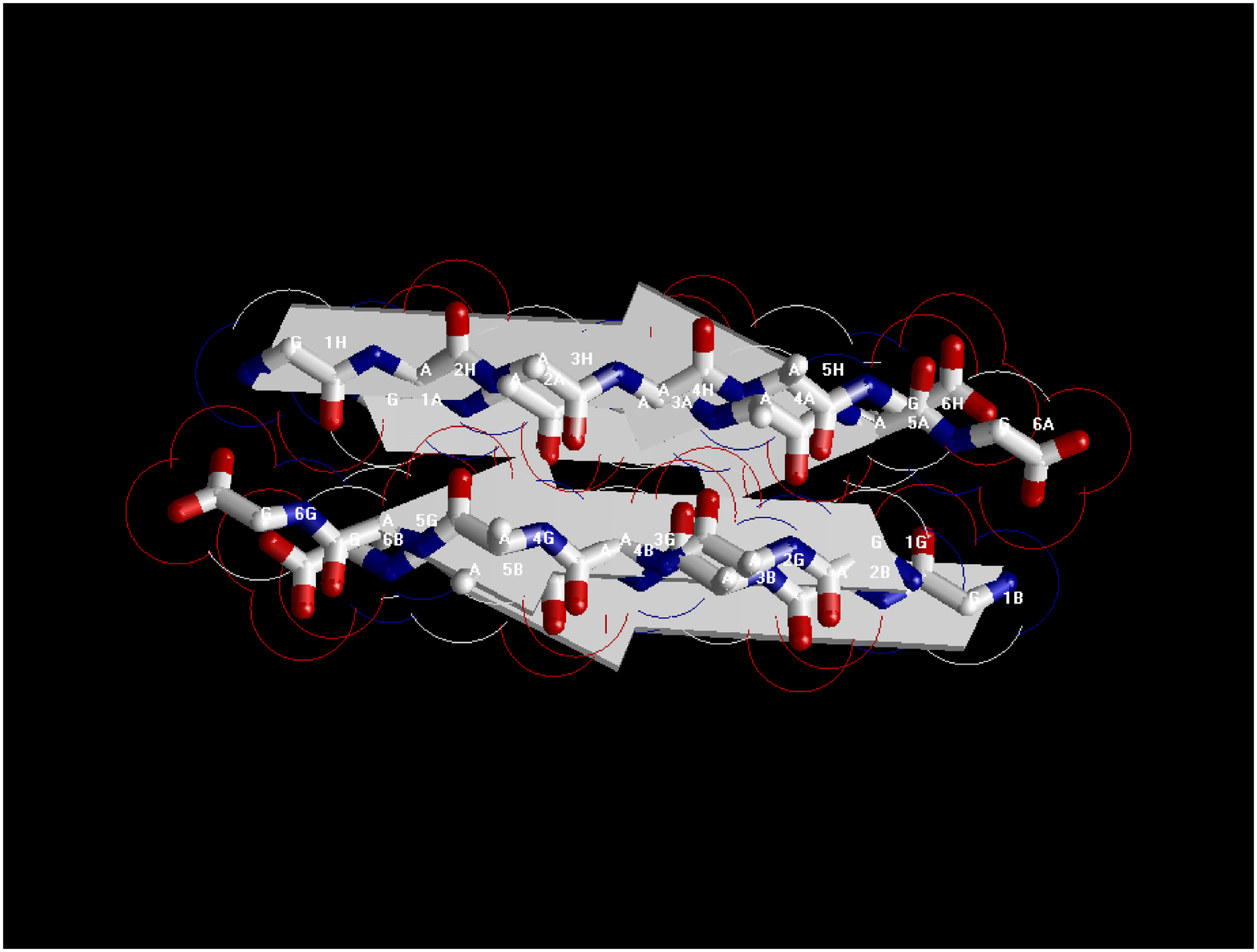}
}
\end{figure*}
\begin{figure}[h!]
\centerline{
\includegraphics[scale=0.6]{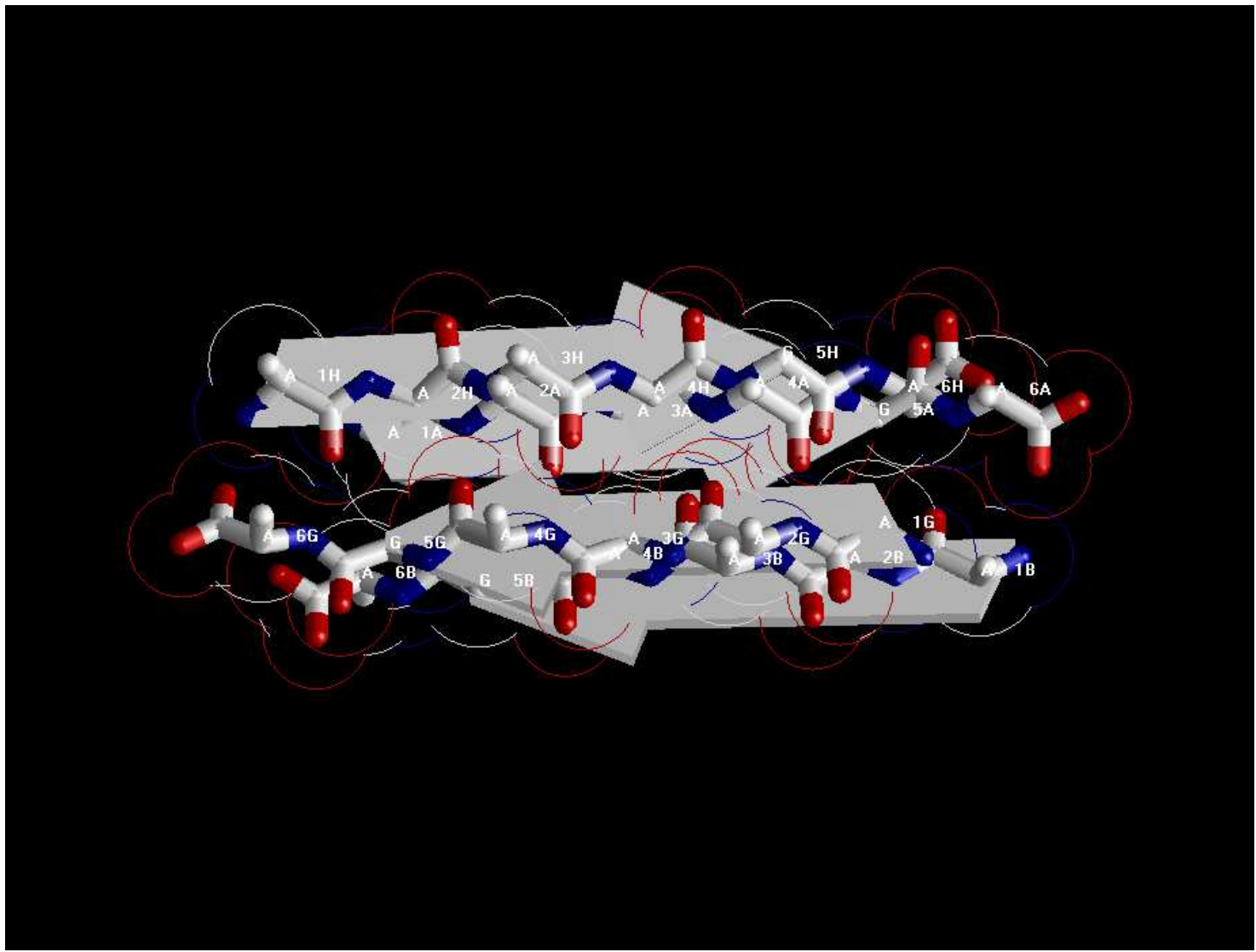}
}
\caption{Close vdw contacts of AG chains and BH chains of Models 1-3.}
\label{fig3}
\end{figure}

\begin{figure*}[h!]
\centerline{
\includegraphics[scale=0.6]{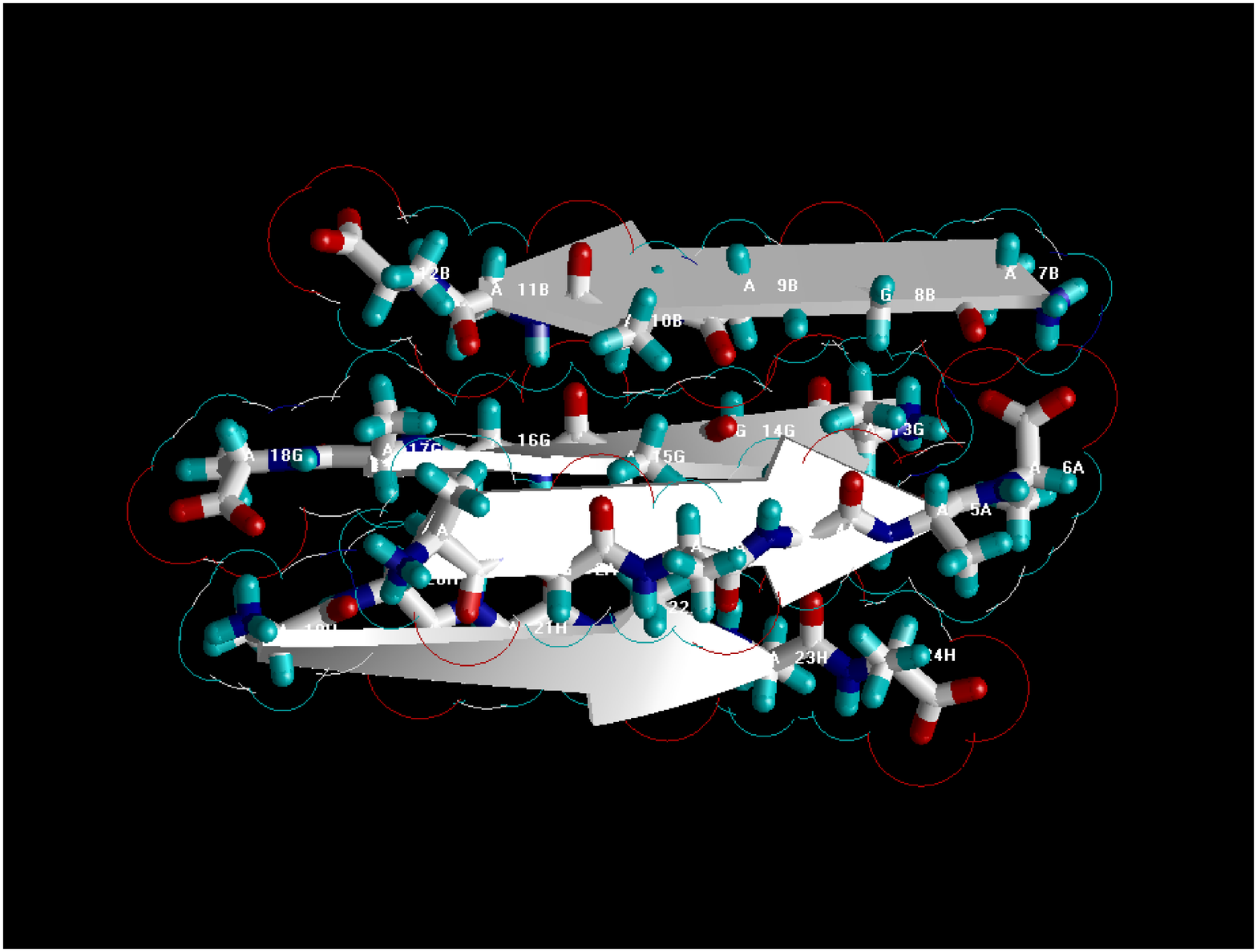}
}
\end{figure*}
\begin{figure*}[h!]
\centerline{
\includegraphics[scale=0.6]{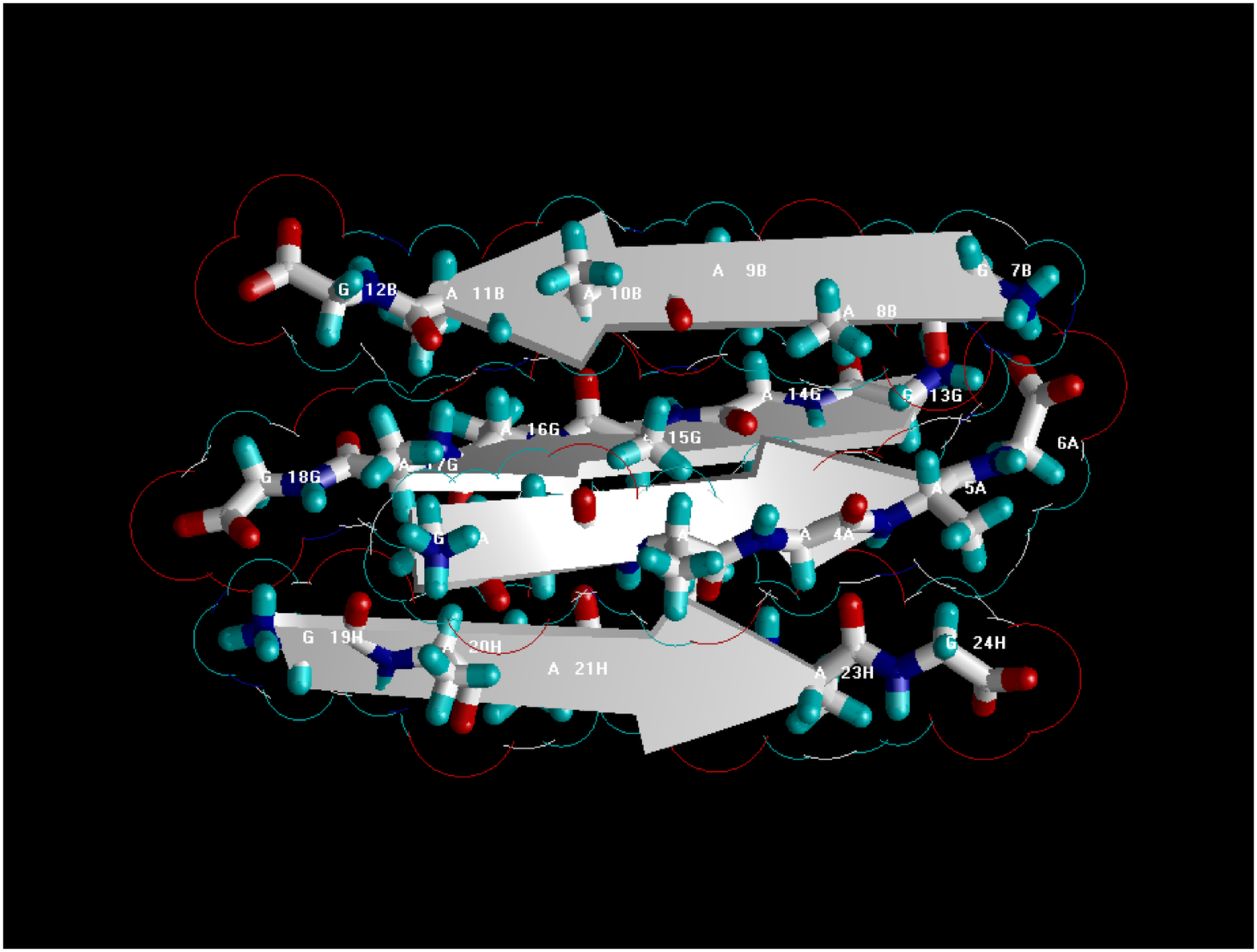}
}
\end{figure*}
\begin{figure}[h!]
\centerline{
\includegraphics[scale=0.6]{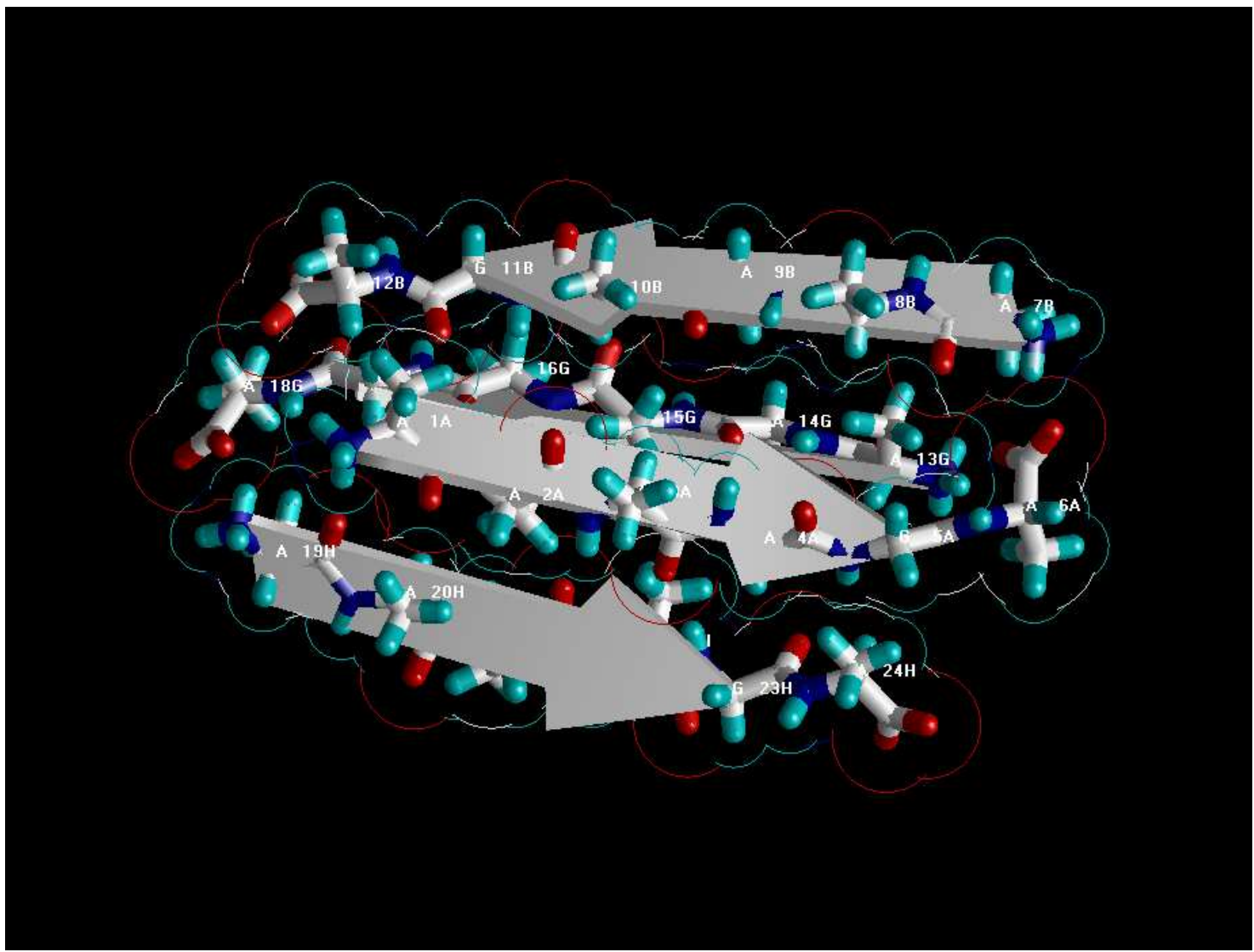}
}
\caption{Optimal structures of prion AAAAGA amyloid fibril Models 1-3.}
\label{fig4}
\end{figure}
\end{document}